\documentclass[12pt]{article}

\usepackage{graphicx}
\usepackage{natbib}
\usepackage{url} 
\usepackage{multirow}%
\usepackage{amsmath,amssymb,amsfonts}%
\usepackage{amsthm}%
\usepackage{mathrsfs}%
\usepackage[title]{appendix}%
\usepackage{xcolor}%
\usepackage{textcomp}%
\usepackage{manyfoot}%
\usepackage{booktabs}%
\usepackage{algorithm}%
\usepackage{algorithmicx}%
\usepackage{algpseudocode}%
\usepackage{listings}%
\usepackage{caption}
\usepackage{subcaption}
\usepackage{bbm}
\usepackage{hyperref}

\DeclareMathOperator*{\argmin}{arg\,min}

\newcommand{\blind}{0}

\definecolor{DarkBlue}{rgb}{0,0,.545}
\hypersetup{
    colorlinks = true,
    linkcolor = DarkBlue,
    filecolor = DarkBlue,      
    urlcolor = DarkBlue,
    citecolor = DarkBlue,
}

\begin{document}

\def\spacingset#1{\renewcommand{\baselinestretch}%
{#1}\small\normalsize} \spacingset{1}


\if0\blind
{
  \title{\bf Adaptive Shrinkage with a Nonparametric Bayesian Lasso}
  \author{Santiago Marin
  \hspace{.2cm}\\
    Australian National University\\
    and \\
    Bronwyn Loong \\
    Australian National University \\
    and \\
    Anton H. Westveld \\
    Australian National University and Virginia Commonwealth University
    }
  \maketitle
} \fi

\if1\blind
{
  \bigskip
  \bigskip
  \bigskip
  \begin{center}
    {\LARGE\bf Title}
\end{center}
  \medskip
} \fi

\bigskip
\begin{abstract}
Modern approaches to perform Bayesian variable selection rely mostly on the use of shrinkage priors. That said, an ideal shrinkage prior should be adaptive to different signal levels, ensuring that small effects are ruled out, while keeping relatively intact the important ones. With this task in mind, we develop the nonparametric Bayesian Lasso, an adaptive and flexible shrinkage prior for Bayesian regression and variable selection, particularly useful when the number of predictors is comparable or larger than the number of available data points. We build on spike-and-slab Lasso ideas and extend them by placing a Dirichlet Process prior on the shrinkage parameters. The result is a prior on the regression coefficients that can be seen as an infinite mixture of Double Exponential densities, all offering different amounts of regularization, ensuring a more adaptive and flexible shrinkage. We also develop an efficient Markov chain Monte Carlo algorithm for posterior inference. Through various simulation exercises and real-world data analyses, we demonstrate that our proposed method leads to a better recovery of the true regression coefficients, a better variable selection, and better out-of-sample predictions, highlighting the benefits of the nonparametric Bayesian Lasso over existing shrinkage priors.
\end{abstract}

\noindent%
\vfill

\newpage
\spacingset{1.75} 
\section{Introduction}
\label{sec:intro}
Modern datasets are made out of a large number of variables, so identifying which predictors are the most important ones---also known as variable selection---is now a crucial and important task in modern Statistics and Machine Learning. To be precise, consider the canonical linear model of the form
\begin{equation}
    \label{eq:linear_reg}
    \boldsymbol{y} = \mathbf{X}\boldsymbol{\beta} + \boldsymbol{\varepsilon},\qquad\boldsymbol{\varepsilon}\sim N_{n}(\mathbf{0}_{n}, \sigma^{2}\mathbf{I}_{n}),
\end{equation}
where $\boldsymbol{y}=(y_1,\dots y_n)'\in\mathbb{R}^{n}$ is a vector of responses, $\mathbf{X}\in\mathbb{R}^{n\times p}$ is a matrix of predictors, $\boldsymbol{\beta}=(\beta_{1},\dots,\beta_{p})'\in\mathbb{R}^{p}$ is a vector of regression coefficients, and $\sigma^{2}>0$ is the sampling variance. Without loss of generality, we assume that the response vector and the design matrix have been centered about zero, so the intercept term is not included in the regression function. The \textit{least absolute shrinkage and selection operator} (Lasso), proposed by \cite{tibshirani1996regression}, aims to perform both regularization and variable selection by minimizing the least-squares loss function with an additional  constraint on the $\ell_{1}$ norm of $\boldsymbol{\beta}$. More formally, the Lasso solution is given by
\begin{equation}
    \label{eq:Lasso_problem}
    \hat{\boldsymbol{\beta}}_{\text{Lasso}} = \argmin_{\boldsymbol{\beta}\in\mathbb{R}^{p}} 
    \left\{\frac{1}{2} \left\Vert \boldsymbol{y} - \mathbf{X}\boldsymbol{\beta} \right\Vert_{2}^{2} + \lambda \left\Vert \boldsymbol{\beta} \right\Vert_{1}\right\},
\end{equation}
where $\lambda$ is a nonnegative tuning parameter controlling the amount of shrinkage. As $\lambda$ increases, the Lasso continuously shrinks the regression coefficients towards zero and, for a sufficiently large $\lambda$, some coefficients will be shrunk to exactly zero.

Note, additionally, that the Lasso solution corresponds to a posterior mode when the regression coefficients have independent and identical Double Exponential priors, all with rate parameter $\lambda$. With this in mind, \cite{park2008bayesian} proposed a full Bayesian implementation of the Lasso, known as the Bayesian Lasso, along with an efficient Markov chain Monte Carlo (MCMC) algorithm for posterior inference.

Nonetheless, as discussed in \cite{CARVALHO_horse} and \cite{Bhadra_lasso_meets_horse}, an ideal regularization method should shrink more aggressively small effects, while leaving relatively intact the important ones. The Lasso and the Bayesian Lasso, however, penalize all the coefficients by the same amount $\lambda$, without differentiating important and less important effects. As a result, the Lasso might end up choosing an incorrect model with a non-vanishing probability, regardless of the sample size and how $\lambda$ is chosen (see e.g., \cite{zou2006adaptive}, \cite{zhao2006model}, and \cite{leng2014bayesian}). In the Bayesian case, \cite{Ghosh_Bayes_risk} showed that the Bayesian Lasso tends to undershrink small coefficients while overshrinking the important ones \citep{bai2021spike}.

To address these issues, \cite{zou2006adaptive} proposed to use a different regularization parameter $\lambda_{j}$ for each regression coefficient $\beta_{j}$, and showed that this procedure, known as the adaptive Lasso, is \textit{oracle} \citep{Fan_Li_oracle}. Full Bayesian implementations of the adaptive Lasso have been presented in \cite{Wang_graphical} and \cite{leng2014bayesian}, among others. One limitation of the adaptive Lasso is that there are $p$ tuning parameters, and selecting an adequate combination of those tuning parameters is a task as difficult and important as accurately estimating the regression coefficients. In a Bayesian framework, however, one can treat each $\lambda_{j}$ as a random variable with a prior distribution. This allows for an efficient estimation of the shrinkage parameters alongside all other model parameters using an MCMC algorithm in a data-driven manner.

Another popular approach for variable selection within a Bayesian framework is the use of \textit{spike-and-slab} priors, in which the regression coefficients are drawn, \textit{a-priori}, from either a point-mass at zero (the ``spike") or from a more spread continuous distribution (the ``slab"). More recently, \cite{SSLasso_2018} proposed the spike-and-slab Lasso, providing a link between the Lasso and spike-and-slab priors. Generally speaking, the spike-and-slab Lasso can be interpreted as a mixture of two Double Exponential distributions with different rate parameters, one playing the role of the spike and the other one playing the role of the slab. 

Motivated by recent developments within the spike-and-slab Lasso literature (see e.g., \cite{bai2021spike} for a more comprehensive review), throughout this article, we aim to extend the spike-and-slab Lasso prior from a two component mixture of Double Exponential densities to an infinite component mixture. To do so, we build on Bayesian nonparametric ideas, specifically on Dirichlet Process Mixtures, due to their versatility and wide applicability. Consequently, we call our proposed shrinkage prior the nonparametric Bayesian Lasso.

The use of Dirichlet Process Mixtures in regression problems is not new (see e.g., \cite{muller1996bayesian}, \cite{shahbaba2009nonlinear}, and \cite{hannah2011dirichlet}, just to name a few). In the context of shrinkage priors and Bayesian variable selection, we can pinpoint the work by \cite{Dunson_shrinkage}, \cite{Das_Dirichlet}, \cite{Quintana_Partition}, \cite{Barcella_2016}, and \cite{Ding_dirichlet}. These existing methods tend to specify a Dirichlet Process prior on the parameters of the sampling model, with the aim of clustering observed units according to patterns in the covariate space \citep{Barcella_review}. Our proposed method, on the other hand, specifies a Dirichlet Process prior on the regularization parameter of a Bayesian Lasso, aiming for a more adaptive and versatile shrinkage through a mixture of Double Exponential distributions. In that sense, one could say that our work is closer to the spike-and-slab Lasso from \cite{SSLasso_2018}.

We show that, relative to more traditional shrinkage priors, our nonparametric Bayesian Lasso leads to a better recovery of the true regression coefficients, a better variable selection, and better out-of-sample predictions. We also derive an efficient MCMC algorithm, which allows for a systematic posterior computation.

The rest of this article is organized as follows. Section \ref{sec:Background} revisits spike-and-slab Lasso as well as Dirichlet Process Mixtures. In Section \ref{sec:BNPLasso}, we formally introduce our nonparametric Bayesian Lasso and derive the necessary algorithms for posterior computation. Simulation exercises are carried out in
Section \ref{sec:Sims}. In section \ref{sec:benchmark_data}, we demonstrate the applicability of our proposed method on real-world data. We
conclude with a discussion in Section \ref{sec:Discussion}.
\section{Background}
\label{sec:Background}
\subsection{Spike-and-Slab Lasso}
\label{subsec:SSLasso}
Following \cite{SSLasso_2018}, a spike-and-slab Lasso prior for the regression coefficients can be formulated as $p(\boldsymbol{\beta}|\pi_{1},\pi_{2}) = \prod_{j=1}^{p}p(\beta_{j}|\pi_{1},\pi_{2})$, with $p(\beta_{j}|\pi_{1},\pi_{2})$ taking the form
\begin{equation}
    \label{eq:SSLasso_prior}
    p(\beta_{j}|\,\pi_{1}, \pi_{2}) = \sum_{k=1}^{2}\pi_{k}\,\psi(\beta_{j}|\,0, \lambda_{k}),
\end{equation}
where $\psi(\cdot\,|\, 0, \lambda)$ denotes the density of a Double Exponential distribution with location zero and rate parameter $\lambda$, $(\pi_{1}, \pi_{2})\sim p(\pi_{1}, \pi_{2})$, and $\sum_{k}\pi_{k}=1$. It is clear from the above specification that the spike-and-slab Lasso is a finite mixture of two Double Exponential distributions, where $\lambda_{1}\gg\lambda_{2}$ such that $\psi(\cdot\,|\,0, \lambda_{1})$ plays the role of the spike and $\psi(\cdot\,|\,0, \lambda_{2})$ plays the role of the slab. Thus, the prior in \eqref{eq:SSLasso_prior} adaptively thresholds small coefficients through the spike, while keeping large effects unaltered with the heavy-tailed slab \citep{SSLasso_2018}, making the spike-and-slab Lasso an appealing variable selection procedure. What is more, letting $\lambda_{1}=\lambda_{2}$ reduces the method to a traditional Bayesian Lasso, while letting $\lambda_{1}\rightarrow\infty$ results in a limiting point-mass spike-and-slab prior, effectively linking both methodologies \citep{bai2021spike}.
\subsection{Dirichlet Process Mixtures}
\label{subsec:DPMs}
Dirichlet Process Mixtures are one of the most popular Bayesian nonparametric (BNP) models. More formally, consider a random probability measure, $H$, completely unrelated to our variable selection problem at hand. Then, following \cite{muller2015bayesian}, one could place a Dirichlet Process (DP) prior on $H$ \citep{Ferguson_DP}. Namely,
\begin{equation*}
    H\sim\text{DP}(\alpha, H_{0}),
\end{equation*}
where $\alpha$ is a concentration parameter and $H_0$ is a centering measure defined over the same space as $H$. That being said, an important property of the DP is that it generates random probability measures that are almost surely (a.s.) discrete. In other words, $H$ is discrete with probability of one, even if $H_0$ is absolutely continuous. Based on the discrete nature of the process, \cite{sethuraman1994constructive} consequently proposed a stick-breaking representation of the form
\begin{equation*}
    H(\cdot) = \sum_{k=1}^{\infty}w_{k}\delta_{H_{k}}(\cdot),
\end{equation*}
where $\delta$ denotes the Dirac measure, $\{H_{k}\}_{k\in\mathbb{N}}\overset{\text{iid}}{\sim}H_{0}$, $w_k = \nu_{k}\prod_{l<k}(1 - \nu_{l})$, and $\{\nu_{k}\}_{k\in\mathbb{N}}\overset{\text{iid}}{\sim}\text{Beta}(1,\alpha)$. 

In many applications, however, the a.s. discreteness of the DP might be too restrictive. To overcome this, one can consider a Dirichlet Process Mixture \citep{Antoniak_DPM}, which is a convolution of $H$ with some continuous parametric kernel. More formally, let $z_{1},\dots,z_{n}$ be some arbitrary random variables unrelated to our variable selection problem at hand, and let $f(z|\theta)$ be a continuous parametric kernel. Then, a mixture of $f(z|\theta)$ with respect to $H$ has a probability density function given by
\begin{equation}
\label{eq:DPConvolution}
f(z|H) = \int f(z|\theta)dH(\theta).
\end{equation}

We can equivalently write the mixture in \eqref{eq:DPConvolution} as a hierarchical model of the form
\begin{gather*}
    z_{i}|\theta_{i}\overset{\text{ind}}{\sim}f(z|\theta_i),\qquad
    \theta_{i}|H\overset{\text{iid}}{\sim}H, \qquad
    H\sim\text{DP}(\alpha, H_0).
\end{gather*}
Moreover, due to the a.s. discreteness of the DP, we have that 
\begin{equation}
    \label{eq:DPInf}
    f(z|H) = \sum_{k=1}^{\infty}w_{k}f(z|\theta_{k}),
\end{equation}
where $\{\theta_{k}\}_{k\in\mathbb{N}}\overset{\text{iid}}{\sim}H_{0}$. If, for example, we set our parametric kernel to be the density of a Double Exponential distribution with location zero and rate parameter $\theta$, we would have that 
\begin{equation}
    \label{eq:DPInf_Lasso}
    f(z|H) = \sum_{k=1}^{\infty}w_{k}\psi(z|\,0, \theta_{k}).
\end{equation}

Note the close similarities between the density in \eqref{eq:DPInf_Lasso} and the spike-and-slab Lasso prior in \eqref{eq:SSLasso_prior}. The main difference is that spike-and-slab Lasso corresponds to a two component mixture of Double Exponential densities, while the density in \eqref{eq:DPInf_Lasso} corresponds to a mixture with a countably infinite amount of components, so it can be seen as a generalization of the spike-and-slab Lasso prior.

\begin{figure}
\begin{center}
\includegraphics[width=\textwidth]{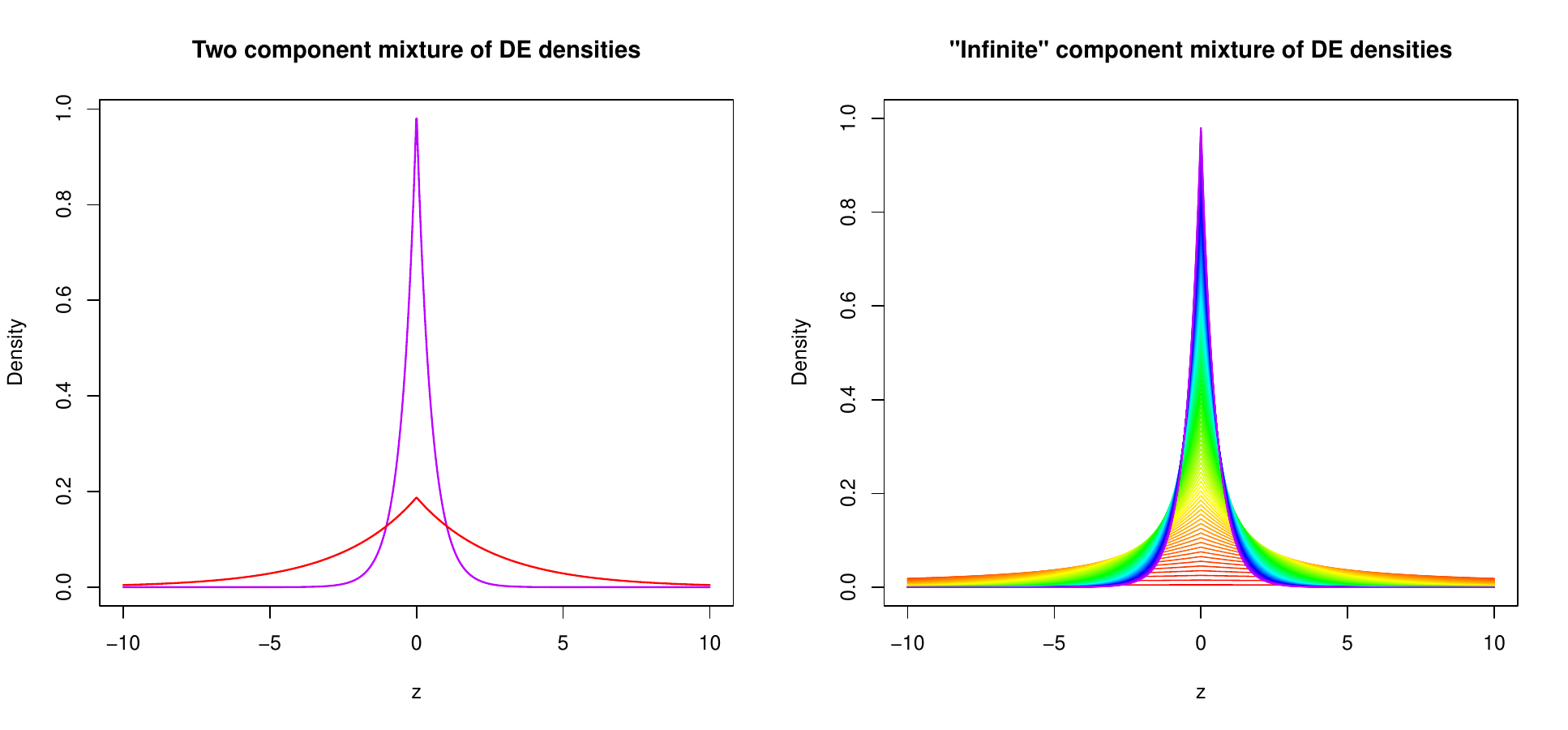}
\end{center}
\caption{A two component mixture and an ``infinite" component mixture of Double Exponential (DE) densities. Violet curves represent Double Exponential densities with a larger rate parameter and red curves represent Double Exponential densities with a smaller rate parameter. The effect of all other rate parameters is depicted according to the colors of the rainbow.
\label{fig:DEdensities}}
\end{figure}

To illustrate these differences more clearly, Figure \ref{fig:DEdensities} presents a two component mixture and an ``infinite" component mixture of Double Exponential densities. In the plot, violet curves represent Double Exponential densities with a larger rate parameter and red curves represent Double Exponential densities with a smaller rate parameter. The effect of all other rate parameters is depicted according to the colors of the rainbow. 

It is clear from Figure \ref{fig:DEdensities} that considering a two component mixture of Double Exponential densities provides an advantage over the Bayesian Lasso as one can adapt the shrinkage strength to different signal levels. In particular, we can see that the spike (depicted in violet) regularizes more aggressively small coefficients, while the heavy-tailed slab (depicted in red) ensures that we do not overshrink large effects \citep{SSLasso_2018}.

Note, however, that the infinite component mixture provides a ``rainbow" of alternative shrinkage parameters, ensuring a more adaptive and flexible regularization. For instance, if the shrinkage parameter corresponding to the violet Double Exponential density results in an excessive shrinkage, but the shrinkage parameter corresponding to the red Double Exponential density applies too little shrinkage, the infinite component mixture could then adapt the shrinkage parameter accordingly, potentially ensuring an adequate amount of regularization for each regression coefficient.

\section{The Nonparametric Bayesian Lasso}
\label{sec:BNPLasso}

\subsection{Model Formulation}
\label{subsec:model}
Following \cite{park2008bayesian}, we make use of the fact that the Double Exponential distribution can be represented as a scale mixture of Normals with an Exponential mixing density \citep{Andrews_scalesNormal}. More precisely, our full model is
\begin{gather*}
    \boldsymbol{y}|\,\mathbf{X},\boldsymbol{\beta},\sigma^{2}\sim N_{n}(\mathbf{X}\boldsymbol{\beta}, \sigma^{2}\mathbf{I}_{n}), \\
    \{\beta_j\}_{j=1}^{p}|\,\{\tau_{j}^{2}\}_{j=1}^{p}\overset{\text{ind}}{\sim}N (0, \tau_{j}^{2}), \\
    p(\{\tau_{j}^{2}\}_{j=1}^{p}|\{\lambda_{j}^{2}\}_{j=1}^{p})\propto\prod_{j=1}^{p}\frac{\lambda_{j}^{2}}{2}\exp\left\{-\frac{\lambda_{j}^{2}}{2}\tau_{j}^{2}\right\}, \\
    \{\lambda_{j}^{2}\}_{j=1}^{p}|\,G \overset{\text{iid}}{\sim} G, \\
    G \sim \text{DP}(\alpha, G_0), 
\end{gather*}
where $\alpha$ is a concentration parameter and $G_0$ is a prior centering measure. To complete our model specification, we set $G_0$ to be a Gamma distribution with shape parameter $a$ and rate parameter $b$. For our prior on the sampling variance, $\sigma^{2}$, we follow \cite{moran2019a} and specify an independent prior of the form $p(\sigma^2)\propto1/\sigma^2$ so that, \textit{a-priori}, $\boldsymbol{\beta}$ and $\sigma^2$ are independent. Finally, we assume that the data are centered, so the intercept has been integrated out from the model.

Following the discussion in section \ref{subsec:DPMs}, our prior formulation corresponds to an infinite mixture of Double Exponential densities---all with different rate parameters---as our prior on $\boldsymbol{\beta}$, effectively nesting both the spike-and-slab Lasso and the Bayesian Lasso as special cases. What is more, by considering a countably infinite amount of shrinkage parameters, our proposed method is able to adapt the amount of shrinkage to different signal levels, regularizing more aggressively small coefficients while keeping large effects relatively intact. Considering the fact that our adaptive shrinkage prior follows from a Dirichlet Process Mixture, we call our proposed method the nonparametric Bayesian Lasso.
\subsection{Posterior Inference}
\label{subsec:post_inference}
Given our specified likelihood and priors, the Bayesian posterior distribution is---up to a proportionality constant---given by
\begin{equation*}
    p(\boldsymbol{\beta}, \boldsymbol{\tau}, \boldsymbol{\lambda}, \sigma^{2} | \boldsymbol{y}, \mathbf{X}) \propto p(\sigma^{2})\prod_{j=1}^{p}p(\beta_j|\tau_{j}^{2},\sigma^2)\int p (\tau_{j}^{2}|\lambda_{j}^{2})dG(\lambda_{j}^2)\prod_{i=1}^{n}p(y_{i}|\mathbf{x}_{i}, \boldsymbol{\beta}, \sigma^{2}).
\end{equation*}

To generate draws from the above posterior distribution, our MCMC algorithm sequentially samples from the full conditional distributions of all the model parameters. Sampling from the full conditional distributions of $\boldsymbol{\beta}$, $\boldsymbol{\tau}$, and $\sigma^2$ is a standard procedure (see e.g., \cite{park2008bayesian}), so we now focus on sampling from the full conditional distribution of each $\lambda_{j}^2$. For the sake of completeness, we present implementation details of the remaining steps in the supplementary materials. 

\textbf{Sampling $\lambda^2$:} To sample each $\lambda_{j}^2$ from its respective full conditional distribution, we make use of the Gibbs sampler proposed in \cite{Maceachern_1994}. Following \cite{muller2015bayesian}, we say that $\tau^2_j$ and $\tau^2_{j'}$, for $j\neq j'$, are clustered together if and only if $\lambda^{2}_{j} = \lambda^{2}_{j'}$. In other words, $\tau^2_j$ and $\tau^2_{j'}$ must share the same shrinkage parameter in order to be clustered together. Thus, let us define $\lambda_{k}^{2^{*}}$ as the common parameter in the $k$-th cluster, for $k\in\{1,\dots,K\}$. Additionally, let us introduce the latent membership indicators $r_1,\dots,r_p$, such that  
\begin{equation*}
    r_{j} = 
    \begin{cases}
        k, \quad \text{iff} \quad \tau_{j}^{2} \quad  \text{belongs to the $k$-th cluster}, \\
        0, \quad \text{iff} \quad \tau_{j}^{2} \quad  \text{is not in any existing clusters}.
    \end{cases}
\end{equation*}

Then, we can sample $\lambda_{k}^{2^{*}}$, for $k\in\{0,\dots,K\}$, from the full conditional distribution
\begin{equation*}
    \begin{split}
        p(\lambda_{k}^{2^{*}}|\,\boldsymbol{r}, \boldsymbol{\tau}) & \propto G_0(d\lambda_{k}^{2^{*}})\prod_{j:r_{j}=k}p(\tau_{j}^{2}|\lambda_{k}^{2^{*}}) \\
        & \propto \left(\lambda_{k}^{2^{*}}\right)^{(a + p_k)-1}\exp\left\{-\left(b + \frac{1}{2}\sum_{j:r_{j}=k}\tau_{j}^{2}\right)\lambda_{k}^{2^{*}} \right\},
    \end{split}
\end{equation*}
which is the kernel of a $\text{Gamma}(a + p_k,\, b + \frac{1}{2}\sum_{j:r_{j}=k}\tau_{j}^{2})$ distribution, where $p_k$ is the size of the $k$-th cluster. What is more, we can then sample each $r_j\sim\text{Multinomial}(\varphi_{j0}, \varphi_{j1}, \dots, \varphi_{jK})$, where $\varphi_{j0}$ is the posterior probability that $\tau_{j}^{2}$ is not in any of the current clusters, and it is given by
\begin{equation*}
    \begin{split}
        \varphi_{j0} & \propto \alpha \int_{\mathbb{R}^{+}} p(\tau_{j}^{2}|\lambda_{j}^{2})G_0(d\lambda_{j}^{2}) \\
        & = \alpha \frac{b^{a}}{2\Gamma(a)} \int_{\mathbb{R}^{+}} \left(\lambda_{j}^{2}\right)^{(a+1)-1}\exp\left\{-\left(b + \frac{\tau_{j}^{2}}{2}\right)\lambda_{j}^{2} \right\} d\lambda_{j}^{2} \\
        & = \alpha \frac{ab^{a}}{2\left(b + \frac{\tau_{j}^{2}}{2}\right)^{(a+1)}}.
    \end{split}
\end{equation*}
Moreover, for $k\in\{1,\dots,K\}$, $\varphi_{jk}$ is the posterior probability that $\tau_{j}^{2}$ belongs to the $k$-th cluster, and it is given by
\begin{equation*}
    \begin{split}
        \varphi_{jk} & \propto p_{k,-j} \times p(\tau_{j}^{2}|\lambda_{j}^{2}) \\
        & \propto p_{k,-j} \times \frac{\lambda_{j}^{2}}{2}\exp\left\{-\frac{\lambda_{j}^{2}}{2}\tau_{j}^{2}\right\},
    \end{split}
\end{equation*}
where $p_{k,-j}$ is the size of the $k$-th cluster, excluding $\tau_j^{2}$. 

Our MCMC algorithm, then, keeps sampling from the full conditional distributions of all the model parameters. The implementation details of the remaining steps are given in the supplementary materials. 

Unlike similar approaches, like the ones proposed in \cite{Dunson_shrinkage} or \cite{Das_Dirichlet}, our nonparametric Bayesian Lasso results in a relatively simple Gibbs sampler, allowing for more efficient computations. Moreover, our DP prior over the shrinkage parameters yields a prior on the regression coefficients that can be seen as an infinite mixture of Double Exponential densities, in the spirit of the spike-and-slab Lasso, illustrating the novelty of our proposed method.

\subsection{Variable Selection}
\label{subsec:var_select}
A well-known property of continuous shrinkage priors like the Bayesian Lasso, the Bayesian adaptive Lasso, or the nonparametric Bayesian Lasso, is that the posterior distribution of the regression coefficients is absolutely continuous. In other words, the regression coefficients do not have a positive probability of being exactly zero. Therefore, in order to perform variable selection, we follow the scaled neighborhood criterion from \cite{Li_Lin_elastic}. More precisely, we estimate each $\beta_{j}$, for $j\in\{1,\dots,p\}$, as
\begin{equation}
    \label{eq:var_select}
    \hat{\beta}_{j} = \begin{cases}
        0,\qquad\qquad\quad \text{if  } \; \mathbb{P}(\beta_j \in B_{j}|\boldsymbol{y}, \mathbf{X}) > 0.5,\\
        \mathbb{E}[\beta_j|\boldsymbol{y}, \mathbf{X}], \quad \text{otherwise},
    \end{cases}
\end{equation}
where $\mathbb{P}(\beta_j \in B_{j}|\boldsymbol{y}, \mathbf{X})$ is the marginal posterior probability that $\beta_{j}$ is in the scaled neighborhood ${B}_{j} = [-\sqrt{\mathbb{V}(\beta_j|\boldsymbol{y}, \mathbf{X})},\, \sqrt{\mathbb{V}(\beta_j|\boldsymbol{y}, \mathbf{X})} ]$, while $\mathbb{E}[\beta_j|\boldsymbol{y}, \mathbf{X}]$ and $\mathbb{V}(\beta_j|\boldsymbol{y}, \mathbf{X})$ are the marginal posterior mean and variance of each $\beta_j$, respectively.

\section{Numerical Simulations}
\label{sec:Sims}
We now conduct simulation exercises to assess the performance of our nonparametric Bayesian Lasso.
More precisely, we consider different sample sizes $n\in\{100,\, 250,\, 500\}$ and a fixed dimension $p=200$. For the design matrix, $\mathbf{X} = (\mathbf{x}_{1},\dots,\mathbf{x}_{n})'\in\mathbb{R}^{n\times p}$, we generate each covariate vector as $\mathbf{x}_{i}\sim N_{p}(\mathbf{0}_{p}, \boldsymbol{\Sigma})$ with $\Sigma_{i,j} = \left\{\rho^{|i-j|}\right\}_{i,j=1}^{p}$ and $\rho\in\{0.3,\, 0.5,\, 0.7\}$ to induce different correlations between the predictors. The true coefficient vector, $\boldsymbol{\beta}_{\text{true}}$, is constructed as
\begin{equation*}
    \boldsymbol{\beta}_{\text{true}} = (\,\underbrace{10, \dots, 10}_{5}, \,\underbrace{2, \dots, 2}_{15}, \,\underbrace{0, \dots, 0}_{180}\,)'\in\mathbb{R}^{p},
\end{equation*}
so that only 10\% of the entries are nonzero. However, there are few (five) very strong signals, while there are more (fifteen) weaker signals. The response vector, conditional on $\mathbf{X}$ and $\boldsymbol{\beta}_{\text{true}}$, is then generated as $\boldsymbol{y}\sim N_{n}(\mathbf{X}\boldsymbol{\beta}_{\text{true}}, \mathbf{I}_{n})$. 
To evaluate the model fit, we also simulate held-out data of size $n_{\text{test}} = 1000$, denoted as $\{{y}_{\text{test}, i}, \mathbf{x}_{\text{test}, i}\}_{i=1}^{n_{\text{test}}}$.  

For each sample size $n$, we simulate $L=200$ different datasets and consider the following traditional assessment metrics: 
\begin{itemize}
    \item $\text{MSE} = \frac{1}{L}\sum_{l=1}^{L}\left(\frac{1}{p}\sum_{j=1}^{p}\left(\beta_{\text{true}, j} - \hat{\beta}_{j}^{(l)}\right)^{2}\right)$,
    \item $\text{Selection Accuracy} = \frac{1}{L}\sum_{l=1}^{L}\left(\frac{1}{p}\sum_{j=1}^{p}\mathbbm{1}\left\{\text{sign}(\beta_{\text{true}, j}) = \text{sign}(\hat{\beta}_{j}^{(l)}) \right\}\right)$, 
    \item $\text{MSPE} = \frac{1}{L}\sum_{l=1}^{L}\left(\frac{1}{n_{\text{test}}}\sum_{i=1}^{n_{\text{test}}} \left({y}_{\text{test}, i}^{(l)} - \hat{y}_{\text{test}, i}^{(l)} \right)^{2}\right)$,
\end{itemize}
where $\beta_{\text{true}, j}$ denotes the $j$-th entry of the true regression coefficient vector, $\hat{\beta}_{j}^{(l)}$ is its corresponding estimate based on the $l$-th training dataset, ${y}_{\text{test}, i}^{(l)}$ is the $i$-th response in the $l$-th held-out set, and $\hat{y}_{\text{test}, i}^{(l)}$ is its corresponding prediction based on the $l$-th training dataset. Hence, MSE and selection accuracy evaluate how well we are recovering the true regression coefficients and how well we are performing variable selection, respectively, while MSPE assess the out-of-sample predictive performance. In all cases, $\hat{\boldsymbol{\beta}}^{(l)}$ is constructed as in \eqref{eq:var_select}.

Note, however, that these metrics are all based on point estimates and do not take into account any uncertainty quantification. Since we have access to a sequence of posterior draws, we can make use of the expected log pointwise predictive density (elppd) as an additional assessment metric, which takes into account posterior uncertainty. In particular, maximizing the elppd is equivalent to a better fit on the held-out set (see e.g., \cite{gelman2013bayesian}). Given a held-out set, $\{{y}_{\text{test}, i}, \mathbf{x}_{\text{test}, i}\}_{i=1}^{n_{\text{test}}}$, we compute the elppd as
\begin{equation*}
    \text{elppd}=\frac{1}{n_{\text{test}}}\sum_{i=1}^{n_{\text{test}}}\log\left(\frac{1}{S}\sum_{s=1}^{S}p\left({y}_{\text{test}, i} | \mathbf{x}_{\text{test}, i}^{_{'}}\boldsymbol{\beta}_{(s)}, \sigma^{2}_{(s)}\right)\right),
\end{equation*}
where $\boldsymbol{\beta}_{(s)}$ and $\sigma^{2}_{(s)}$ are the $s$-th draws of $\boldsymbol{\beta}$ and $\sigma^{2}$ in our MCMC algorithm, and $S$ is the total number of draws. Then, we can compute a mean elppd as $\frac{1}{L}\sum_{l=1}^{L}\text{elppd}^{(l)}$, where $\text{elppd}^{(l)}$ is the elppd for the $l$-th dataset. 

We compare our proposed method to the Bayesian Lasso and the Bayesian adaptive Lasso. For our prior centering measure, $G_0$, we consider a Gamma distribution with shape parameter $a=0.1$ and rate parameter $b=0.1$, and we set our concentration parameter to be $\alpha=0.01$. Similarly, for the Bayesian Lasso, we assume that the regularization parameter is drawn, \textit{a-priori}, from a Gamma distribution also with shape parameter $a=0.1$ and rate parameter $b=0.1$, and for the Bayesian adaptive Lasso, we place independent and identical Gamma priors, all with shape $a=0.1$ and rate $b=0.1$, one each regularization parameter. The Bayesian Lasso and the Bayesian adaptive Lasso are implemented in the Stan probabilistic programming language \citep{carpenter_stan}. In all cases, we obtain 6000 posterior draws and discard the first 1000 draws as \textit{burn-in}.
\subsection{Simulation Results}
\label{subsec:sim_res}
Detailed simulation results are presented in Table \ref{tab:sim_res}. We can observe that, across all sample sizes and correlation levels, our nonparametric Bayesian Lasso constantly yields the smallest MSE and MSPE (where smaller values suggest a better performance), while yielding the largest variable selection accuracy and elppd (where larger values suggest a better performance). These differences are more clear under larger correlations and smaller sample sizes. For instance, when $n=100$ and $\rho=0.7$, the MSE from the nonparametric Bayesian Lasso is 5.5 times smaller than the one from the Bayesian Lasso and 4.3 times smaller than the one from the Bayesian adaptive Lasso. Moreover, the MSPE from our proposed method is 3.74 and 2.09 times smaller than the ones from the Bayesian Lasso and the Bayesian adaptive Lasso, respectively. We can also observe that our proposed method provides the best variable selection and the largest elppd.


These results are consistent across a wide variety of simulation settings, illustrating the benefits of our shrinkage prior over more traditional Bayesian methods. Altogether, our proposed methodology provides a better recovery of the true regression coefficients, a better variable selection, and better out-of-sample predictions.  

%
%

\begin{table}[h]
    \caption{simulation results for the nonparametric Bayesian Lasso (BNP-L), the Bayesian Lasso (B-L), and the Bayesian adaptive Lasso (BA-L).}\label{tab:sim_res}
    \begin{tabular*}{\textwidth}{llccccccccc}
    \toprule%
    & & \multicolumn{3}{@{}c@{}}{\small{$n=100$}} & \multicolumn{3}{@{}c@{}}{\small{$n=250$}} & \multicolumn{3}{@{}c@{}}{\small{$n=500$}} \\
    \cmidrule(lr){3-5} 
    \cmidrule(lr){6-8}
    \cmidrule(lr){9-11}
    & & \footnotesize{BNP-L} & \footnotesize{B-L} & \footnotesize{BA-L} & \footnotesize{BNP-L} & \footnotesize{B-L} & \footnotesize{BA-L} & \footnotesize{BNP-L} & \footnotesize{B-L} & \footnotesize{BA-L} \\
    \midrule
    \footnotesize{$\rho=0.3$} & \footnotesize{MSE} & \footnotesize{\textbf{0.003}} & \footnotesize{0.068} & \footnotesize{0.011} & \footnotesize{\textbf{0.001}} & \footnotesize{0.008} & \footnotesize{0.004} & \footnotesize{\textbf{0.001}} & \footnotesize{0.003} & \footnotesize{0.002} \\
    & \footnotesize{Sel. Acc.} & \footnotesize{{0.971}} & \footnotesize{0.970} & \footnotesize{\textbf{0.981}} & \footnotesize{\textbf{0.931}} & \footnotesize{0.748} & \footnotesize{0.841}  & \footnotesize{\textbf{0.893}} & \footnotesize{0.718} & \footnotesize{0.797} \\
    & \footnotesize{MSPE} & \footnotesize{\textbf{1.553}}  & \footnotesize{18.093} & \footnotesize{2.929}  & \footnotesize{\textbf{1.232}} & \footnotesize{2.463} & \footnotesize{1.752} & \footnotesize{\textbf{1.137}} & \footnotesize{1.465} & \footnotesize{1.314} \\
    & \footnotesize{Mean elppd} & \footnotesize{\textbf{-1.761}}  & \footnotesize{-3.294}  & \footnotesize{-2.588}  & \footnotesize{\textbf{-1.581}} & \footnotesize{-1.934} & \footnotesize{-1.774} & \footnotesize{\textbf{-1.512}} & \footnotesize{-1.638} & \footnotesize{-1.588} \\
    \midrule
    \footnotesize{$\rho=0.5$} & \footnotesize{MSE} & \footnotesize{\textbf{0.004}} & \footnotesize{0.039} & \footnotesize{0.011} & \footnotesize{\textbf{0.001}} & \footnotesize{0.010} & \footnotesize{0.005} & \footnotesize{\textbf{0.001}} & \footnotesize{0.003} & \footnotesize{0.002} \\
    & \footnotesize{Sel. Acc.} & \footnotesize{{0.978}} & \footnotesize{0.978} & \footnotesize{\textbf{0.987}} & \footnotesize{\textbf{0.949}} & \footnotesize{0.760} & \footnotesize{0.860}  & \footnotesize{\textbf{0.916}} & \footnotesize{0.727} & \footnotesize{0.816} \\
    & \footnotesize{MSPE} & \footnotesize{\textbf{1.533}}  & \footnotesize{9.682} & \footnotesize{2.396}  & \footnotesize{\textbf{1.209}} & \footnotesize{2.444} & \footnotesize{1.732} & \footnotesize{\textbf{1.125}} & \footnotesize{1.479} & \footnotesize{1.315} \\
    & \footnotesize{Mean elppd} & \footnotesize{\textbf{-1.766}}  & \footnotesize{-3.119}  & \footnotesize{-2.493}  & \footnotesize{\textbf{-1.576}} & \footnotesize{-1.914} & \footnotesize{-1.760} & \footnotesize{\textbf{-1.508}} & \footnotesize{-1.633} & \footnotesize{-1.583} \\
    \midrule
    \footnotesize{$\rho=0.7$} & \footnotesize{MSE} & \footnotesize{\textbf{0.006}} & \footnotesize{0.033} & \footnotesize{0.026} & \footnotesize{\textbf{0.002}} & \footnotesize{0.014} & \footnotesize{0.007} & \footnotesize{\textbf{0.001}} & \footnotesize{0.005} & \footnotesize{0.003} \\
    & \footnotesize{Sel. Acc.} & \footnotesize{\textbf{0.991}} & \footnotesize{0.968} & \footnotesize{0.978} & \footnotesize{\textbf{0.972}} & \footnotesize{0.779} & \footnotesize{0.887}  & \footnotesize{\textbf{0.947}} & \footnotesize{0.742} & \footnotesize{0.845} \\
    & \footnotesize{MSPE} & \footnotesize{\textbf{1.464}}  & \footnotesize{5.478} & \footnotesize{3.073}  & \footnotesize{\textbf{1.176}} & \footnotesize{2.474} & \footnotesize{1.753} & \footnotesize{\textbf{1.110}} & \footnotesize{1.529} & \footnotesize{1.337} \\
    & \footnotesize{Mean elppd} & \footnotesize{\textbf{-1.752}}  & \footnotesize{-2.906}  & \footnotesize{-2.475}  & \footnotesize{\textbf{-1.564}} & \footnotesize{-1.877} & \footnotesize{-1.737} & \footnotesize{\textbf{-1.500}} & \footnotesize{-1.624} & \footnotesize{-1.573} \\
    \toprule%
    \end{tabular*}
        
    \vspace{0.1cm}
    
    \small{Note: The best performance, across each sample size and correlation $\rho$, is presented in bold.}
\end{table}

\section{Protein Activity Data}
\label{sec:benchmark_data}
To further demonstrate the practical utility of our proposed method, we now apply the Bayesian Lasso, the Bayesian adaptive Lasso, and the nonparametric Bayesian Lasso to the protein activity data from \cite{clyde1998protein} and \cite{clyde2011bayesian}. The data consists of $n=96$ observations and $p=88$ potential predictors. The aim is to identify which of these 88 features have a large predictive power of ``protein activity level during storage."

As discussed in \cite{clyde1996prediction} and \cite{clyde2011bayesian}, this dataset impose a significant challenge to traditional shrinkage priors due to the large colinearities between predictors. More precisely, 17 pairs of variables have correlations above 0.95 and the maximum correlation between predictors is 0.99. What is more, we expect that only a handful of the 88 features will have large predictive power.

\begin{figure}
\begin{center}
\includegraphics[width=0.85\textwidth]{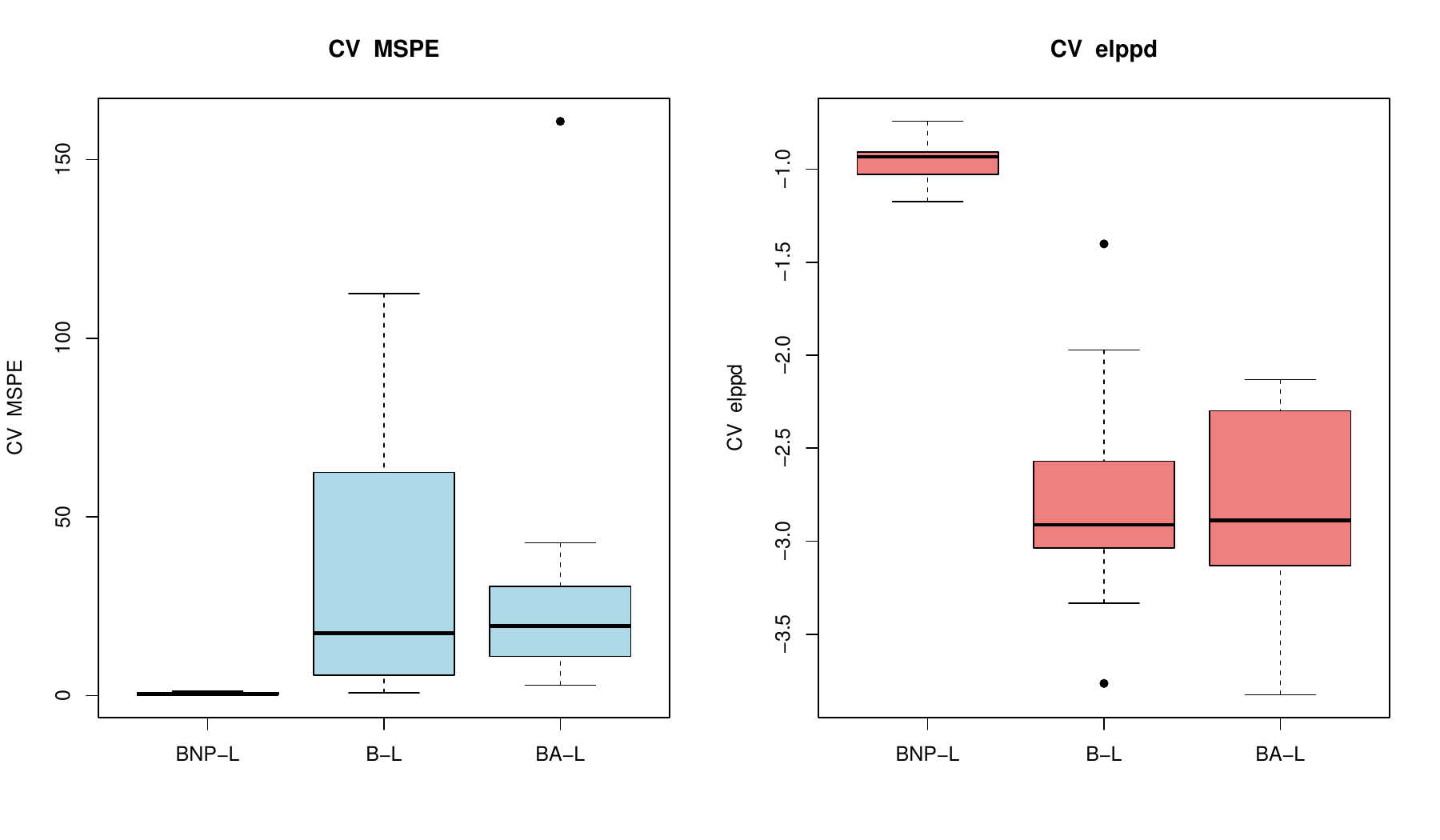}
\end{center}
\caption{Boxplots of a ten-fold cross validation experiment on the protein activity data for the nonparametric Bayesian Lasso (BNP-L), the Bayesian Lasso (B-L), and the Bayesian adaptive Lasso (BA-L). \label{fig:Box_protein}}
\end{figure}

We evaluate the predictive performance of each one of the three competing methods trough ten-fold cross validation. As comparison metrics we use the cross validation MSPE and the cross validation elppd. Figure \ref{fig:Box_protein} and Table \ref{tab:benchmark_res} present detailed results of our ten-fold cross validation experiment.

\begin{table}[h]
    \caption{Results of a ten-fold cross validation experiment on the protein activity data for the nonparametric Bayesian Lasso (BNP-L), the Bayesian Lasso (B-L), and the Bayesian adaptive Lasso (BA-L).}\label{tab:benchmark_res}
    \begin{tabular*}{\textwidth}{@{\extracolsep\fill}lccc}
    \toprule%
    & BNP-L & B-L & BA-L  \\
    \midrule
    Average CV MSPE & \textbf{0.619} & 34.394 & 33.464 \\
    Average CV elppd & \textbf{-0.955} & -2.765 & -2.843 \\
    \toprule%
    \end{tabular*}
    
    \vspace{0.1cm}
    
    \small{Note: The best performance is presented in bold.}
\end{table}

We can observe that our proposed method produces the smallest cross validation MSPE (where smaller values suggest a better prediction) and the largest cross validation elppd (where larger values suggest a better fit to the test data). These results are encouraging as the prediction errors for the other competing methods are significantly larger, illustrating the superior performance of our shrinkage prior. More precisely, note that the average cross validation MSPE from our proposed method is 55.56 times smaller than the one from the Bayesian Lasso and 54.06 times smaller than the one from the Bayesian adaptive Lasso. This illustrates the advantages of considering a more flexible and adaptive shrinkage prior like the one proposed in this article.

Additionally, to evaluate the variable selection accuracy from each method, we follow \cite{moran2019a} and compare the results from the nonparametric Bayesian Lasso, the Bayesian Lasso, and the Bayesian adaptive Lasso, against the Bayesian adaptive sampling (BAS) algorithm \citep{clyde2011bayesian}, which has been effectively utilized on the protein data and is available on the \texttt{R} package \texttt{BAS} \citep{BAS_pckg}. The BAS algorithm found seven important predictors, namely: 
\begin{itemize}
    \item \texttt{con}: The protein concentration,
    \item \texttt{detN}: Detergent N,
    \item \texttt{detT}: Detergent T,
    \item \texttt{bufTRS:detN}: The interaction between buffer TRS and detergent N,
    \item \texttt{bufPO4:temp}: The interaction between buffer P04 and temperature,
    \item \texttt{con:detT}: The interaction between protein concentration and detergent T,
    \item \texttt{detN:temp}: The interaction between Detergent N and temperature.
\end{itemize}
The Bayesian Lasso and and the Bayesian adaptive Lasso found 23 and 56 important predictors, respectively, failing to provide any meaningful variable selection. The nonparametric Bayesian Lasso, on the other hand, found six important predictors, namely:
\begin{itemize}
    \item \texttt{con}: The protein concentration,
    \item \texttt{bufTRS:con}: The interaction between buffer TRS and protein concentration,
    \item \texttt{bufPO4:temp}: The interaction between buffer P04 and temperature,
    \item \texttt{NaCl:con}: The interaction between NaCl and protein concentration,
    \item \texttt{NaCl:detT}: The interaction between NaCl and detergent T,
    \item \texttt{con:detT}: The interaction between protein concentration and detergent T.
\end{itemize}
Out of the six predictors identified by the nonparametric Bayesian Lasso, three were also selected by BAS. Note, additionally, that the correlation between \texttt{detT} (identified as an important predictor by BAS) and \texttt{NaCl:detT} (identified as an important predictor by the nonparametric Bayesian Lasso) is 0.74, implying a strong colinearity between the two variables. This could be a reason on why one algorithm is picking one variable and not the other. On the whole, the results of the nonparametric Bayesian Lasso are partially consistent with the ones from previous studies.

\begin{figure}
\begin{center}
\includegraphics[width=\textwidth]{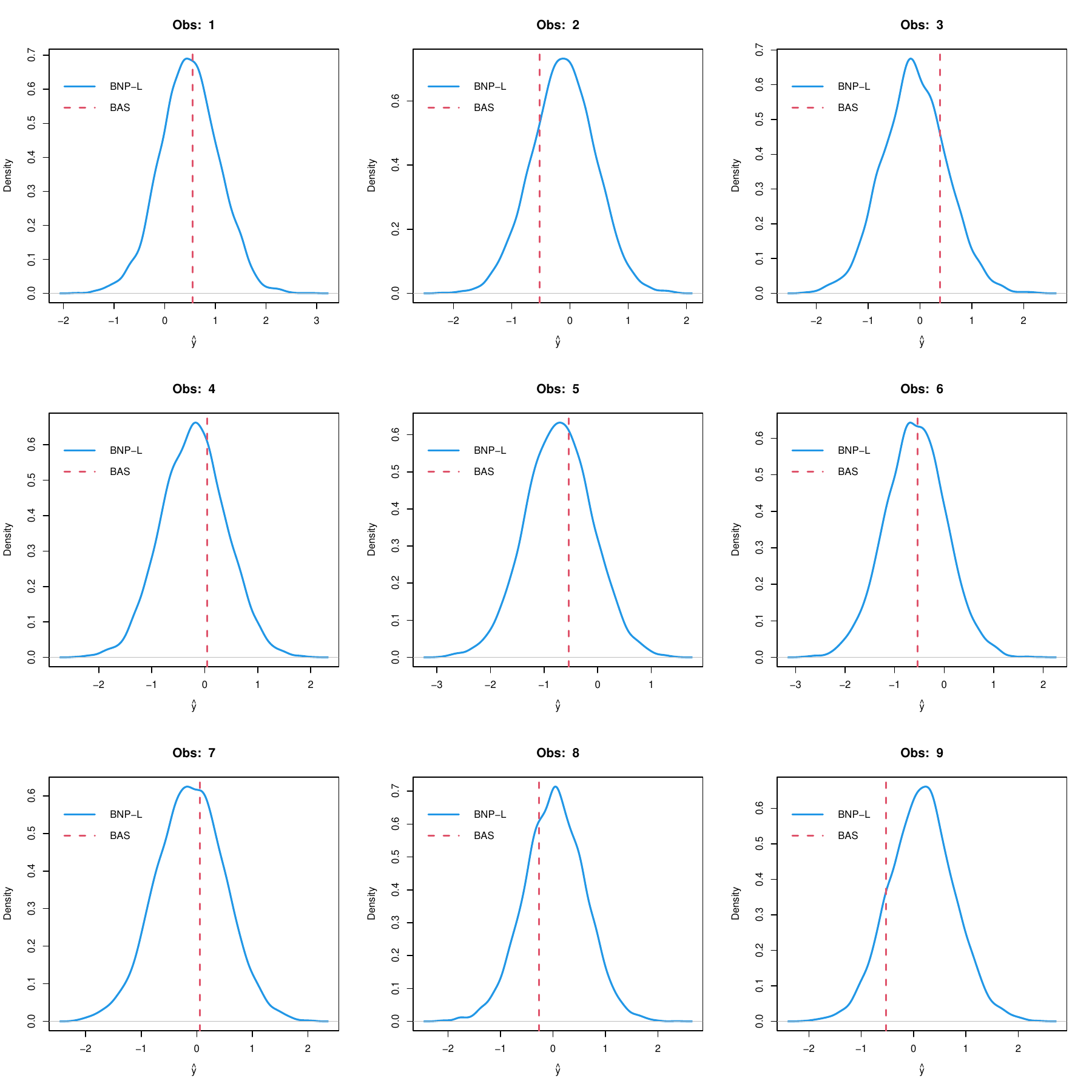}
\end{center}
\caption{Posterior predictive distribution of the fitted values for the first nine observations in the protein dataset obtained with the nonparametric Bayesian Lasso (the blue densities), along with their corresponding fitted values obtained with BAS (the red dashed lines). \label{fig:Post_preds}}
\end{figure}

Lastly, we would like to compare the posterior predictive distribution of the fitted values obtained with our nonparametric Bayesian Lasso to the fitted values obtained with BAS. Figure \ref{fig:Post_preds} present the posterior predictive distribution of the fitted values for the first nine observations in the dataset obtained with the nonparametric Bayesian Lasso (the blue densities), along with their corresponding fitted values obtained with BAS (the red dashed lines). 

It is clear from Figure \ref{fig:Post_preds} that most of the fitted values obtained with BAS align with the posterior predictive distribution of the fitted values obtained with our nonparametric Bayesian Lasso, suggesting close similarities between the two methods. For a more formal analysis, we also examine if the fitted values obtained with BAS are located in areas of large posterior predictive density. More precisely, let $p_{\hat{y}_{i}^{(\text{BNP-L})}}(\cdot)$ be the posterior predictive density of the fitted value for observation $i$ obtained with the nonparametric Bayesian Lasso, and let $\hat{y}_{i}^{(\text{BAS})}$ be the corresponding fitted value obtained with BAS. Then, we can consider the following quantity as a comparison metric,
\begin{equation*}
    \text{Average Fitted Density}^{(\text{BNP-L})} = \frac{1}{n}\sum_{i=1}^{n} p_{\hat{y}_{i}^{(\text{BNP-L})}}\left( \hat{y}_{i}^{(\text{BAS})} \right).
\end{equation*}

If the posterior predictive distribution of the fitted values obtained with the nonparametric Bayesian Lasso is symmetric and centered about the fitted values obtained with BAS (which is desirable), the above average fitted density should be large, suggesting that the BAS solution is in an area of large posterior predictive density. On the other hand, smaller values would suggest larger discrepancies between the two methods. Note, additionally, that we can also compute an $\text{Average Fitted Density}^{(\text{B-L})}$ and an $\text{Average Fitted Density}^{(\text{BA-L})}$, which are the average fitted densities based on the Bayesian Lasso and the Bayesian adaptive Lasso, respectively.

Table \ref{tab:fitted_dens} presents the average fitted densities for the nonparametric Bayesian Lasso, the Bayesian Lasso, and the Bayesian adaptive Lasso. We can observe that the nonparametric Bayesian Lasso yields the largest average fitted density, suggesting closer similarities between BAS and our proposed method.

\begin{table}[h]
    \caption{Average fitted densities for the nonparametric Bayesian Lasso (BNP-L), the Bayesian Lasso (B-L), and the Bayesian adaptive Lasso (BA-L).}\label{tab:fitted_dens}
    \begin{tabular*}{\textwidth}{@{\extracolsep\fill}lccc}
    \toprule%
    & BNP-L & B-L & BA-L  \\
    \midrule
    Average Fitted Density & \textbf{0.575} & 0.151 & 0.068 \\
    \toprule%
    \end{tabular*}
    
    \vspace{0.1cm}
    
    \small{Note: The best performance is presented in bold.}
\end{table}

Overall, these results suggest that the nonparametric Bayesian Lasso could be applied in challenging real-world data analyses, where traditional shrinkage priors tend to struggle more.
\section{Discussion}
\label{sec:Discussion}
In this article, we have developed the nonparametric Bayesian Lasso, an adaptive and flexible shrinkage prior for Bayesian regression and variable selection, particularly useful when the number of predictors is comparable or larger than the number of available data points. We discussed its connections to state-of-art shrinkage priors, like the spike-and-slab Lasso, as well as with Bayesian nonparametric ideas, particularly with Dirichlet Process Mixtures and infinite component mixtures. We also developed an efficient MCMC algorithm for posterior inference. Through various simulation exercises and real-world data analyses, we have highlighted the benefits of our nonparametric Bayesian Lasso over existing shrinkage priors. More precisely, our proposed method provides a better recovery of the true regression coefficients, a better variable selection, and better out-of-sample predictions, making the nonparametric Bayesian Lasso a competitive shrinkage prior.

One concern, however, is with respect to the scalability of our nonparametric Bayesian Lasso to ultra high-dimensional settings, where one observes hundreds of thousands of covariates. This is because our posterior sampler relies on traditional MCMC algorithms, which work very efficiently when the number of predictors is moderately large. Nonetheless, due to the sequential nature of MCMC algorithms and the cost of the linear algebra operations associated with them, they usually do not scale well to ultra high-dimensional settings (see e.g., \cite{Bayes_boot_SSLASSO} or \cite{marin2023bob} for a discussion). Alternative implementations based on Variational Bayes \citep{Blei_variational} could be explored in the future. 

On the whole, our nonparametric Bayesian Lasso joins the Probabilistic Machine Learning ``toolbox", expanding and enriching the already vibrant world of Bayesian shrinkage priors.

\bibliographystyle{chicago}
\bibliography{citations.bib}

\newpage

\begin{center}
    \LARGE{\bf{Supplementary Materials for ``Adaptive Shrinkage with a Nonparametric Bayesian Lasso"}}
    \\
    \vspace{50px}
    \large{Santiago Marin \qquad Bronwyn Loong \qquad Anton H. Westveld}
\end{center}

\vspace{15px}

\section*{S.1 MCMC Algorithm}
\label{sec:MCMC_details}
Given the specified likelihood and priors, we can simulate from the joint posterior distribution by sequentially sampling from the full conditional distributions of all the model parameters. In the main document, we showed how to sample from the full conditional distribution of each $\lambda_{j}^{2}$, for $j\in\{1,\dots,p\}$. Here we will show how to sample from the full conditional distributions of $\boldsymbol{\beta}$, $\boldsymbol{\tau}$, and $\sigma^{2}$.

\textbf{Sample $\boldsymbol{\beta}$:} Since the likelihood and the conditional prior for $\boldsymbol{\beta}$ are multivariate Gaussian, the full conditional distribution of $\boldsymbol{\beta}$ is also given by a $p$-dimensional multivariate Gaussian distribution of the form 
\begin{gather*}
    \boldsymbol{\beta}|\boldsymbol{y}, \mathbf{X}, \boldsymbol{\tau},\sigma^{2} \sim N_{p}(\boldsymbol{\Phi}^{-1}\mathbf{X}'\boldsymbol{y}/\sigma^{2},\, \boldsymbol{\Phi}^{-1}),\\
    \boldsymbol{\Phi} = \mathbf{X}'\mathbf{X}/\sigma^{2} + \text{diag}(\tau_{1}^{2}, \dots\tau_{p}^{2})^{-1}.
\end{gather*}

\textbf{Sample $\boldsymbol{\tau}$:} Let us define $\gamma_{j}=1/\tau^{2}_{j}$, for $j\in\{1,\dots,p\}$. Then, following \cite{park2008bayesian}, the full conditional distribution of each $\gamma_{j}$ is given by
\begin{equation*}
    p(\gamma_{j}|\beta_j, \lambda_{j}^{2}) = \sqrt{\frac{\Tilde{\lambda}_{j}}{2\pi}}\gamma_{j}^{-3/2}\exp\left\{-\frac{\Tilde{\lambda}_{j}(\gamma_j - \Tilde{\xi}_{j})^{2}}{2\Tilde{\xi}_{j}\gamma_{j}} \right\},
\end{equation*}
which is the kernel of an inverse-Gaussian distribution with parameters
\begin{equation*}
    \Tilde{\xi}_{j} = \sqrt{\frac{\lambda_{j}^{2}}{\beta_{j}^{2}}}, \quad \text{and} \quad \Tilde{\lambda}_{j} = \lambda_{j}^{2}.
\end{equation*}

\textbf{Sample $\sigma^{2}$:} Lastly, the full conditional distribution of $\sigma^{2}$ is an inverse-Gamma distribution with shape parameter $(n-1)/2$ and scale parameter
\begin{equation*}
    \frac{\left\Vert\boldsymbol{y} - \mathbf{X}\boldsymbol{\beta}\right\Vert_{2}^{2}}{2}.
\end{equation*}

\end{document}